# Investigation of Self-similar Properties of Additive Data Traffic

Igor Ivanisenko, Lyudmyla Kirichenko, Tamara Radivilova

*Abstract* - **The work presents results of numerical study of self-similar properties of additive data traffic. It is shown that the value of Hurst exponent of total stream is determined by the maximum value of Hurst exponent of summed streams and the ratio of variation coefficient of stream with maximum Hurst exponent and other ones.**

## I. INTRODUCTION

Numerous researches of processes in a network have shown that statistical characteristics of the traffic have property of time scale invariance (self-similarity). Self-similar properties were discovered in the local and global networks, particularly traffic Ethernet, ATM, applications TCP, IP, VoIP and video streams. The reason for this effect lies in the features of the distribution of files on servers, their sizes, and the typical behavior of users. It was found that initially not having self-similarity data streams passing on nodal processing servers and active network elements became self-similar.

The self-similar traffic has the special structure that preserves on many scales. There are always a number of extremely large bursts at relatively small average level of the traffic. These bursts are cause significant delays and losses of packages, even when the total load of all streams are more less than maximal values. In a classical case for Poisson stream buffers of an average size will be enough. The queue can be formed in short-term prospect, but for the long period buffers will be cleared. However in a case of self-similar traffic queues have more greater length.

For the majority of networks is actual that without restrictions of the incoming traffic, queue on the most loaded lines will grow without limit, and eventually will exceed the sizes of buffers in corresponding units. This can cause the situation, when incoming packets will be ignored and thus will have to be transmitted again, that leads to irrational expenditure of network resources.

Thus, an important task to improve the network quality of service is the study of the properties of self-similar traffic. A characteristic feature of computer networks is the multiplexing of streams, so characteristics of the additive self-similar streams have special significance. In [1-2], theoretical and numerical properties of self-similar additive processes were studied. It was shown that sum of several self-similar processes with different values of the Hurst exponent has maximal one. In [3-5], the results of experimental studies of the properties of additive data traffic which confirm the theoretical results are presented. However, these studies do not take into account traffic burstability, the quantitative characteristic of which is the coefficient of variation.

The purpose of the present work is the numerical investigation of the properties of the additive modeling self-similar traffic, which have varying degrees of burstability.

## II. SELF-SIMILARITY OF NETWORK TRAFFIC

Stochastic process $X(t)$ is statistically self-similar with self-similarity parameter $H$, if the process $a^{-H}X(at)$ has the same statistical properties of the second order as $X(t)$. Parameter $H$, $0 < H < 1$, named Hurst exponent, is the measure of self-similarity and long-term dependence of the stochastic process. The value $H = 0.5$ denotes the absence of long-term dependence.

One of the most important properties of traffic as a random process is the presence of heavy tails of its distribution function. The heaviness of the distribution tails corresponds to the degree of burstability. The coefficient of variation can be considered as the easiest quantitative characteristic of the distribution tail:

$$\sigma_{\text{var}}(T) = \frac{\sigma(T)}{E(T)} \qquad (1)$$

where $T$ is random variable whose values are the numbers of events in the time interval $T$.

## IV. MULTIPLEXING OF SELF-SIMILAR STREAMS

Modern information networks are built on the multiplex data streams. Consider the mechanism of statistical multiplexing of information streams, which is widely used in telecommunications, because it allows to economical use of the bandwidth of the main channels. It consists in the fact that the individual sources are added streams in the main channel with saving bandwidth. Assuming independence and absence of long-term dependence of streams coefficient of variation of the resulting process in the main channel will

Igor Ivanisenko - Kharkiv National University of Radio Electronics, Lenin Avenue 14, Kharkiv, 61166, UKRAINE,
E-mail: ihorek@kture.kharkiv.ua

Lyudmyla Kirichenko - Kharkiv National University of Radio Electronics Lenin Avenue 14, Kharkiv, 61166, UKRAINE,
E-mail: ludmila.kirichenko@gmail.com

Tamara Radivilova- Kharkiv National University of Radio Electronics, Lenin Avenue 14, Kharkiv, 61166, UKRAINE,
E-mail: tamara.radivilova@gmail.com

decrease and the resulting process will be much smoother.

However, if at least one of the streams is self-similar, the total stream acquires the property of self-similarity [1]. If some self-similar streams with the different Hurst exponents are summed, the resulting stream has the a maximum one. In this case, the total stream is not smoothed and statistical multiplexing algorithm is inefficient.

## V. MODEL OF SELF-SIMILAR TRAFFIC

The main tool for the study and predict the behavior of self-similar data streams is simulation, which requires a model of self-similar input load. There are many models of self-similar traffic. In [6] proposed a model of aggregated self-similar stream, taking into account the degree of self-similarity and the "heavy tails" of the distribution function. The model parameters are the intensity of traffic, the Hurst exponent and the coefficient of variation, that corresponds to burstability in the realizations.

The modeling traffic realization is
$$Y(t) = b \cdot \mathrm{Exp}[k \cdot X(t)], \quad (2)$$
where $X(t)$, $t=1,...,N$ is the series of fractal Gaussian noise with the Hurst exponent $H$; $N$ is length of series; $b$ and $k$ are the parameters that regulate the frequency and magnitude of the bursts.

The stochastic process $Y(t)$ is a self-similar stochastic process as the same Hurst exponent $H$, as the initial fractal Gaussian noise. The variable $Y(t)$ has a log-normal distribution.

## V. RESEARCH RESULTS

In the work the investigation of self-similar properties of total streams of various types was carried out. Each of the modeling realizations was based on the transformation (2).

Consider the sum of two streams: the self-similar one $Y_1(t)$ and one with independent values $Y_2(t)$. Grap of typical realizations is represented in Fig. 1. At the top of the Fig. 1 the realization of a self-similar stream with the theoretical Hurst exponent $H_1 = 0.8$ and the coefficient of variation $\sigma 1_{var} = 1.2$ is shown. In the middle part of the Fig. 1 the realization of traffic, which has independent values is shown. The theoretical Hurst exponent $H_2 = 0.5$ and the coefficient of variation $\sigma 1_{var} = 1.2$. The bottom plot shows the realization of the total stream $Y_\Sigma(t)$.

Estimating of Hurst exponent for the total stream showed that when relation $\sigma 1_{var} \approx \sigma 2_{var}$ holds the value of Hurst exponent $H_\Sigma$ on average equals the Hurst exponent $H_1$.

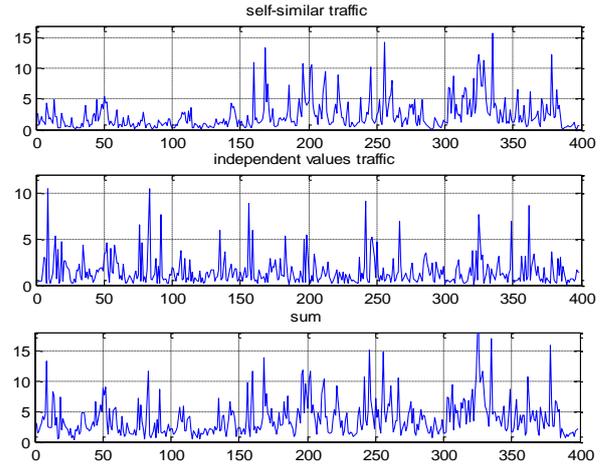

Fig. 1. Model realizations: self-similar, independent values and total

Similar investigations were conducted for the case where the stream $Y_2(t)$ was the short-term dependence process. In this case, the forming process $X(t)$ in the transformation (2) was autoregressive process $X(t) = \varphi X(t-1) + \varepsilon(t)$, where $\varphi$ – autoregression coefficient, $\varepsilon(t)$ – white noise. When relation $\sigma 1_{var} \approx \sigma 2_{var}$ is true the value of the Hurst exponent $H_\Sigma$ on average equals the Hurst exponent $H_1$ also.

Thus, the additive process of self-similar and not self-similar processes acquires the property of self-similarity, when the variation coefficients of streams are close enough. Similar results are given in [4].

Now increase the variation coefficient $\sigma 2_{var}$ of not self-similar stream $Y_2(t)$. In the case when the ratio
$$R_1 = \frac{\sigma 1_{var}}{\sigma 2_{var}}$$
becomes much less than unity, Hurst exponent of total process $H_\Sigma$ gradually decreases, reaching the value 0.5.

Table 1 shows the estimates of the Hurst exponent for the self-similar stream $\hat{H}_1$ (theoretical value of the Hurst exponent $H_1 = 0.8$, the length of realization is 1000) and one of total stream $\hat{H}_\Sigma$, depending on the ratio $R_1$. In this case, $Y_2(t)$ was the stream with independent values.

TABLE 1

PARAMETERS OF SUMMED AND TOTAL REALIZATIONS

| $\hat{H}_1$ | 0.802 | 0.788 | 0.812 | 0.801 | 0.795 |
|---|---|---|---|---|---|
| $R_1$ | 1 | 0.85 | 0.65 | 0.5 | 0.35 |
| $\hat{H}_\Sigma$ | 0.792 | 0.754 | 0.632 | 0.578 | 0.497 |

Now consider the additive stream in the case where two self-similar process are summed. In [4,5] is shown, that the total process $Y_\Sigma(t)$ of the two self-similar processes $Y_1(t)$ and $Y_2(t)$ with Hurst exponents $H_1$ and $H_2$ is self-similar with Hurst exponent $H_\Sigma = \max(H_1, H_2)$. Research was shown that this expression is satisfied, when the processes have the ratio $R_1 \approx 1$. If you increase the value $\sigma 2_{var}$, the Hurst exponent of total stream $H_\Sigma$ will tend to value $H_2$.

Fig. 2 shows the realization of the process $Y_1(t)$ with $H_1 = 0.8$ (dot) and realization of the process $Y_2(t)$ with $H_2 = 0.6$ (solid line). Ratio $R_1 = 0.65$. The additive realization $Y_\Sigma(t)$ has the estimate of Hurst exponent $\hat{H}_\Sigma = 0.714$.

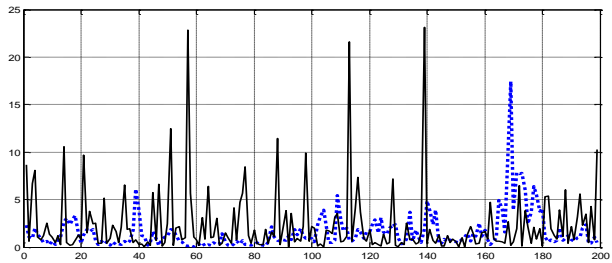

Fig.2. Realizations of summed streams

Table 2 shows the estimates of the Hurst exponents for the self-similar streams $\hat{H}_1$ and $\hat{H}_2$ (theoretical values $H_1 = 0.8$ and $H_2 = 0.6$, the length of realization is 1000) and $H_\Sigma$, depending on the ratio $R_1$.

TABLE 2

PARAMETERS OF SUMMED AND TOTAL REALIZATIONS

| $\hat{H}_1$ | 0.789 | 0.791 | 0.809 | 0.781 | 0.805 |
|---|---|---|---|---|---|
| $\hat{H}_2$ | 0.612 | 0.594 | 0.621 | 0.609 | 0.587 |
| $R_1$ | 1 | 0.85 | 0.65 | 0.5 | 0.35 |
| $\hat{H}_\Sigma$ | 0.805 | 0.732 | 0.714 | 0.634 | 0.612 |

In [3,5] is shown, that the total process of several self-similar processes with equal variances and Hurst exponents $H_i$ is self-similar for which Hurst exponent equals to the maximum Hurst exponent of summed streams: $H_\Sigma = \max(H_i, i = 1,...,N)$. In the summation of several self-similar streams with different coefficients of variation, it is expedient to introduce the coefficient

$$R_2 = \frac{\sigma_{var}(H\max)}{\frac{1}{N-1}\sum_{i=1}^{N-1}\sigma_{var}(i)},$$

where $N$ is number of summed streams, $\sigma_{var}(H\max)$ is the variation coefficient of stream with the greatest Hurst exponent, $\sigma_{var}(i)$ is variation coefficient of $i$-th stream.

Research has shown that in this case the Hurst exponent $H_\Sigma$ of total process $Y_\Sigma(t)$ depends on how the value $R_2$ is less than unity. For values $R_2 \approx 1$ the Hurst exponent $H_\Sigma$ coincides with the maximum value $H_i$.

## V. CONCLUSION

The work presents results of a numerical study of self-similar properties of the additive data streams. The traffic model realizations based on the exponential transformation of fractal Gaussian noise were used as under study data. It has been shown that it is necessary to take into account the ratio of the variation coefficients of summed streams when considering the total stream. The value of the Hurst exponent of the total stream is determined by the maximum value of the Hurst exponent of summed streams and the ratio of coefficient of variation of stream with maximum Hurst and other ones.